\documentclass[stef,twoside]{stefano}

\usepackage{amsmath}
\usepackage{amssymb}
\usepackage{graphics}
\usepackage{rotating}
\usepackage{cite}
\usepackage{color}

\def\makeheadbox{{%
\hbox to0pt{\vbox{\baselineskip=10dd\hrule\hbox
to\hsize{\vrule\kern3pt\vbox{\kern3pt \hbox{{\sc Physical Review D
{\bf 78}, 025006-7 (2008)} }
\hbox{
{\sc {\color{blue}{dma}}[{\color{black}{imecc}}]{\color{red}{UniCamp}}
}
\hspace*{10.3cm}
{\color{blue}{$\boldsymbol{\Sigma \delta \Lambda}$ }}}
\kern3pt}\hfil\kern3pt\vrule}\hrule}%
\hss}}}

\def\0{\mbox{\tiny $0$}}
\def\1{\mbox{\tiny $1$}}
\def\2{\mbox{\tiny $2$}}
\def\3{\mbox{\tiny $3$}}
\def\4{\mbox{\tiny $4$}}
\def\5{\mbox{\tiny $5$}}
\def\6{\mbox{\tiny $6$}}
\def\7{\mbox{\tiny $7$}}
\def\8{\mbox{\tiny $8$}}
\def\9{\mbox{\tiny $9$}}
\def\min1{\mbox{\tiny $(-\,1)$}}
\def\m2{\mbox{\tiny $(-\,2)$}}
\def\={\mbox{\tiny $=$}}

\def\eff{\mbox{\tiny eff}}

\def\NR{\mbox{\tiny NR}}
\def\E{\mbox{\tiny el}}
\def\S{\mbox{\tiny sc}}

\def\Y{\mbox{\tiny Y}}
\def\H{\mbox{\tiny H}}
\def\mi{\mbox{\tiny $-$}}

\begin{document}
%

\title{\Large FERMION-FERMION BOUND STATE CONDITION FOR SCALAR EXCHANGES}

\author{
Stefano De Leo \inst{1}
\and Pietro Rotelli\inst{2} }

\institute{
Department of Applied Mathematics, University of Campinas\\
PO Box 6065, SP 13083-970, Campinas, Brazil\\
{\em deleo@ime.unicamp.br}\\
 \and
Department of Physics, University of
Lecce and INFN Lecce\\
PO BOX 193, CAP 73100, Lecce, Italy\\
{\em rotelli@le.infn.it} }


\date{Submitted {\em April, 2008}}

\abstract{The condition for the existence of a bound state between
two fermions exchanging massive scalars is derived. For low scalar
mass, we reproduce the scalar field model result. The high scalar
mass result exhibits a somewhat different inequality condition.}















\titlerunning{\sc bound states in quantum field theory}

\maketitle


\section*{I. INTRODUCTION}

Since the experimental discovery of the mass of the
neutrinos\cite{PDG} a legitimate question has been posed. Is there
a possibility of a bound state between weakly interacting
particles such as an electron and a neutrino? If the particles
involved where nonrelativistic the answer would be immediate and
negative. From numerical studies of the Schr\"odinger
equation\cite{COHEN} the existence of a bound state produced by a
Yukawa (attractive) coupling
\begin{equation}
V_{\Y}(r)=-\,\frac{g_{\eff}^{\2}}{4\,\pi}\, \frac{\exp[-\,\mu
\,r]}{r} \,\,,
\end{equation}
has been found to be
\begin{equation}
\label{condg} \frac{g^{\2}_{\eff}}{4\,\pi} \ge 0.84\,
\frac{\mu}{m}\,\,,
\end{equation}
where $\mu$ is the exchanged particle mass and $m$ is the reduced
mass. A related derivation is the use of a surrogate to the Yukawa
potential, the Hulthen potential, $V_{\H}$, which approximates the
Yukawa potential for small $r$,
\begin{equation}
V_{\H}(r)=-\,\frac{g_{\eff}^{\2}}{4\,\pi}\,
\frac{2\,\mu}{\exp[2\,\mu\, r] -1}\,\,.
\end{equation}
Our choice of $V_{\H}$ is made so that the terms $r^{\mi \1}$ and
$r^{\0}$ in a series development about $r=0$ are identical to the
Yukawa potential. The Schr\"odinger equation with the Hulthen
potential can be solved {\em analytically}\cite{FLU} and the
existence of a bound state yields a similar result to that above,
i.e.
\begin{equation}
\label{hc}
 \frac{g^{\2}_{\eff}}{4\,\pi} \ge
\frac{\mu}{m}\,\,.
\end{equation}
High mass exchanges would necessitate extremely strong couplings,
obviously unphysical for weak interactions where $\mu/ m > 10^{\1
\1}$\cite{ZUB}. However, if one considers the relativistic
corrections to the Schr\"odinger equation one encounters the well
known Darwin term\cite{SAK}
\begin{equation}
\frac{1}{8\,m}\,\nabla^{^{2}}V(\boldsymbol{r})\,\,,
\end{equation}
which for a Yukawa potential $V_{\Y}(r)$ yields
\begin{equation}
\frac{1}{8\,m^{\2}}\,\left[\,\mu^{\2}\,V_{\Y}(r) +
4\,\pi\,g^{\2}_{\eff}\,\delta(\boldsymbol{r})\,\right]\,\,.
\end{equation}
The first term above can be summed with the potential contribution
to yield an overall amplification factor
\begin{equation}
\left(\,1 + \,\frac{\mu^{\2}}{8\,m^{\2}}\,\right)\,V_{\Y}(r)\,\,.
\end{equation}
This is what has been called Yukawa coupling
amplification\cite{AYC}. Here the effect must be small to comply
with the very nature of correction terms. However, if one where so
bold as to assume this amplification for high $\mu/m$ one would
invert the resonance condition (\ref{hc}), i.e.
\begin{equation}
 \frac{g^{\2}_{\eff}}{4\,\pi} \ge
\frac{8\,m}{\mu}\,\,,
\end{equation}
which allows bound states even for the weak interactions. The
problem, theoretically, now shifts to determining the resonance
condition for high mass exchanges in a more rigorous manner. A
method has been introduced and applied in field
theory\cite{G69,GROSS}. It consists of confronting the lowest
ladder contributions (box and crossed) to the scattering amplitude
at rest, with the tree diagram contribution (also in the rest
frame). The requirement that the sum of the former be equal or
greater than the tree contribution reproduces exactly the Hulthen
condition for low $\mu/m$ within a scalar-scalar model with scalar
particle exchanges. In this paper it will also be shown to be also
valid in the case of fermion-fermion (f-f) interacting  with
scalar exchanges. More recently\cite{EPJC}, the scalar model
calculation was extended to the high $\mu/m$ limit (in either
limit, approximations or numerical calculations are needed). The
high $\mu/m$ result was even more restrictive than the Hulthen
inequality (\ref{hc}), i.e. it required even larger
$g^{\2}_{\eff}$, specifically
\begin{equation}
\frac{g^{\2}_{\eff}}{2\,\pi^{^{\2}}}  \geq
\frac{\mu^{\2}}{m_{\1}m_{\2}} \,\mbox{\large /}\,\left(
\,\ln\frac{\mu^{\2}}{m_{\1}m_{\2}} +
\frac{1+\rho^{\2}}{1-\rho^{\2}} \ln \rho\,\right)\,\,,
\end{equation}
with $\rho=m_{\1}/m_{\2}$. However, it was noted that since the
Klein-Gordon equation lacks a Darwin term correction there is no
reason to expect Yukawa amplification. In this paper, we
essentially repeat our low and high $\mu/m$ limits for f-f
interacting via scalar exchange. This case does contain a Darwin
term identical to that of the well known electrostatic case
although some additional corrections also exist.

In the next Section, we illustrate the model and reduce the first
order ladder contributions to a single integral in
$\mbox{d}|\boldsymbol{k}|=\mbox{d}k$. In Section III, we perform
the small $\mu/m$ limit and reproduce the Hulthen inequality
(\ref{hc}). In Section IV, we perform the high $\mu/m$ limit. We
propose a phenomenological expression for the $k$ integral based
upon numerical simulations. In Section V, we draw our conclusions.

\section*{II. THE FERMIONIC MODEL}

In the center of mass system and for forward scattering (see
Fig.\,1), the Feynman rules\cite{ZUB} for the amplitudes of the
box $(\square)$ and crossed box $(\times)$ diagram yield
\begin{eqnarray}
\mathcal{M}^{\square}(\boldsymbol{p})& =&
i\,g_{\1}^{\2}g_{\2}^{\2} \int\frac{\mbox{d}^{\4}k}{(2\pi)^{\4}}\,
\frac{\bar{u}_{\1}^{(r)}(-\boldsymbol{p})\left\{\left[
E_{\1}(\boldsymbol{p})+E_{\2}(\boldsymbol{p})\right]\gamma_{\0}
-k\hspace*{-.2cm}\slash+m_{\1}
\right\}u_{\1}^{(r')}(-\boldsymbol{p})\,\bar{u}_{\2}^{(s)}(\boldsymbol{p})\left(
k\hspace*{-.2cm}\slash + m_{\2}
\right)u_{\2}^{(s')}(\boldsymbol{p})
}{D^{\square}_{\1}(\boldsymbol{p})D_{\2}(\boldsymbol{p})
D^{^{\2}}_{\0}(\boldsymbol{p})}
\nonumber \\
 & \nonumber \\
 & \\
 & \nonumber \\
\mathcal{M}^{\times}(\boldsymbol{p})& =&i\,g_{\1}^{\2}g_{\2}^{\2}
\int\frac{\mbox{d}^{\4}k}{(2\pi)^{\4}}\,
\frac{\bar{u}_{\1}^{(r)}(-\boldsymbol{p})\left\{k\hspace*{-.2cm}\slash
+m_{\1} + \left[
E_{\1}(\boldsymbol{p})-E_{\2}(\boldsymbol{p})\right]\gamma_{\0}
\right\}u_{\1}^{(r')}(-\boldsymbol{p})\,\bar{u}_{\2}^{(s)}(\boldsymbol{p})\left(
k\hspace*{-.2cm}\slash + m_{\2} \right)u_{\2}^{(s')}(\boldsymbol{p})
}{D^{\times}_{\1}(\boldsymbol{p})D_{\2}(\boldsymbol{p})D^{^{\2}}_{\0}(\boldsymbol{p})}
\nonumber
\end{eqnarray}
with
\[ u^{(s)}_{\1,\2}(\boldsymbol{q}) = \sqrt{E_{\1,\2}(\boldsymbol{q})+m_{{\1,\2}}}\,
 \left( \begin{array}{c} \chi_{s}\\ \\
\displaystyle{\frac{\boldsymbol{\sigma}\cdot
\boldsymbol{q}}{E_{\1,\2}(\boldsymbol{q})+m_{\1,\2}}}\,\chi_s\end{array}
\right)\,\,\,\,\,\,\,\mbox{\small $(s=1,2$)}\,\,,\,\,\,\,\,
\chi_{\1}=\left( \begin{array}{c}1\\0
\end{array} \right)\,\,\,,\,\,\,\,\,\,\,
 \chi_{\2}=\left( \begin{array}{c}0\\1 \end{array} \right)\,\,.
\]
The denominators factors are,
\begin{eqnarray}
D_{\1}^{\square}(\boldsymbol{p}) & = & E^{^{\2}}_{\1}(\boldsymbol{k})-
\left[\, k_{\0} - E_{\1}(\boldsymbol{p})-E_{\2}(\boldsymbol{p})
\right]^{^{\2}} -i\epsilon\,\,,
\nonumber \\
D_{\1}^{\times}(\boldsymbol{p}) & = &
E^{^{\2}}_{\1}(\boldsymbol{k})- \left[\, k_{\0} +
E_{\1}(\boldsymbol{p})-E_{\2}(\boldsymbol{p}) \right]^{^{\2}}
-i\epsilon\,\,,
\nonumber \\
D_{\2}(\boldsymbol{p}) & = & E^{^{\2}}_{\2}(\boldsymbol{k})-
k^{\2}_{\0}-i\epsilon\,\,, \nonumber \\
D_{\0}(\boldsymbol{p}) & = &
E^{^{\2}}_{\0}(\boldsymbol{k}-\boldsymbol{p})-\left[\, k_{\0}
-E_{\2}(\boldsymbol{p}) \right]^{^{\2}} -i\epsilon\,\,,
\end{eqnarray}
where
\[
 E_{\1,\2}(\boldsymbol{q})=\sqrt{\boldsymbol{q}^{\2}+m_{\1,\2}^{\2}}\,\,\,,
\,\,\,\,\,E_{\0}(\boldsymbol{q})=\sqrt{\boldsymbol{q}^{\2}+\mu^{\2}}\,\, .
\]
 At threshold ($\boldsymbol{p}\approx
\boldsymbol{0}$),
\begin{eqnarray}
\mathcal{M}^{\square}(\boldsymbol{0})& =&
i\,\left(\,2\,g_{\1}g_{\2}\sqrt{m_{\1}m_{\2}}\,\right)^{\2}\,\delta_{rr'}\,\delta_{ss'}
\int\frac{\mbox{d}^{\4}k}{(2\pi)^{\4}}\,
\frac{(k_{\0}+m_{\2})(2\,m_{\1}+m_{\2} -
k_{\0})}{D^{\square}_{\1}(\boldsymbol{0})D_{\2}(\boldsymbol{0})
D^{^{\2}}_{\0}(\boldsymbol{0})}\, \, ,
\nonumber \\
 & \\
\mathcal{M}^{\times}(\boldsymbol{0}) & =&
i\,\left(\,2\,g_{\1}g_{\2}\sqrt{m_{\1}m_{\2}}\,\right)^{\2}\,\delta_{rr'}\,\delta_{ss'}
\int\frac{\mbox{d}^{\4}k}{(2\pi)^{\4}}\,
\frac{\left(k_{\0}+2\,m_{\1}-m_{\2}\right)\left(k_{\0}+m_{\2}\right)}
{D^{\times}_{\1}(\boldsymbol{0})D_{\2}(\boldsymbol{0})
D^{^{\2}}_{\0}(\boldsymbol{0})}\,\, . \nonumber
\end{eqnarray}
The poles in the lower half complex $k_{\0}$ plane are at
\[
k_{\0,\1}^{\square}  =  E_{\1}(\boldsymbol{k}) + m_{\1}+
m_{\2}\,\,,\,\,\,\,\, k_{\0,\1}^{\times}  = E_{\1}(\boldsymbol{k})
- m_{\1}+ m_{\2}\,\,, \,\,\,\,\, k_{\0,\2}  =
E_{\2}(\boldsymbol{k}) \,\,, \,\,\,\,\, k_{\0,\0}  =
E_{\0}(\boldsymbol{k}) + m_{\2}\,\,.
\]
The box and crossed box diagrams give the following fourth-order
contribution to the invariant scattering amplitude
\begin{eqnarray}
\label{integral}
\mathcal{M}^{\square}(\boldsymbol{0})+\,\mathcal{M}^{\times}(\boldsymbol{0})
&=&
\frac{\left(\,2\,g_{\1}g_{\2}\sqrt{m_{\1}m_{\2}}\,\right)^{\2}}{(2\pi)^{^{\3}}}\,
\delta_{rr'}\,\delta_{ss'} \int\,\mbox{d}^{\3}k \, \sum_{s=\0}^{\2}
\left[\,
R_{s}^{\square}(\boldsymbol{k})+R_{s}^{\times}(\boldsymbol{k})\,\right] \nonumber \\
 &  = &
\left(\,\frac{g_{\1}g_{\2}\sqrt{2\,m_{\1}m_{\2}}}{\pi}\,\right)^{\2}\,
\delta_{rr'}\,\delta_{ss'} \int_{\0}^{\infty}\mbox{d}k
\,\,k^{\2}\,\,\left[\, R^{\square}(k)+R^{\times}(k)\,\right]\,\,.
\end{eqnarray}
 Below by $E_s$ we intend $E_s(k)$ and by $W$
and $\Delta$ we intend $m_{\1}+m_{\2}$ and $m_{\2}-m_{\1}$
respectively. A simple calculation shows that the explicit
formulas for the residues in the $k_{\0}$-plane for the box and
the crossed box diagram are respectively
\begin{eqnarray}
\label{residues}
 R_{\1}^{\square}(k)
 & = & \left[\,2\,m_{\2}(m_{\1}-E_{\1})-k^{\2} \right]\,/\,\left\{\,4\, W\,E_{\1}
 \left(E_{\1}+m_{\1}\right)\left[\,\mu^{\2}-2\,m_{\1}\left(E_{\1}+m_{\1}\right)\,
\right]^{\2}\right\}\,\,, \nonumber \\
R_{\2}^{\square}(k)
 & = &\left[k^{\2} - 2\, m_{\1}(E_{\2}+m_{\2})\right]\,/\,\left\{\,4\, W\,E_{\2}
 \left(E_{\2}-m_{\2}\right)\left[\,\mu^{\2}+2\,m_{\2}\left(E_{\2}-m_{\2}\right)\,
\right]^{\2}\right\}\,\,,\nonumber \\
R_{\0}^{\square}(k) & = &\left[ 2\,(E_{\0}+2\,
m_{\2})(2\,m_{\1}-E_{\0})\right]
\left[\left(E_{\0}-m_{\1}\right)\,B\,C +
\left(E_{\0}+m_{\2}\right)\,A_{\square}\,C -
\,A_{\square}\,B\,\right]\,/\,
\left[\,A_{\square}^{\2}B^{\2}C^{\3}\,\right] \nonumber \\
 & &
+\, 2\,\left(m_{\1}- m_{\2} -
E_{\0}\right)\,/\,\left[\,A_{\square}\,B\,C^{\2}\,\right]\,\,,
\end{eqnarray}
with $A_{\square}=2\,E_{\0}\,m_{\1} - \mu^{\2}$, $B=-
2\,E_{\0}\,m_{\2} - \mu^{\2}$ and $C=2\,E_{\0}$, and
\begin{eqnarray}
R_{\1}^{\times} & = & \left[\, k^{\2}+2\,m_{\2}
\left(E_{\1}+m_{\1} \right) \right]\,/\,\left\{\,4\,
\Delta\,E_{\1}
 \left(E_{\1}-m_{\1}\right)\left[\,\mu^{\2}+2\,m_{\1}\left(E_{\1}-m_{\1}\right)\,
\right]^{\2} \right\}\,\,,\nonumber \\
R_{\2}^{\times}
 & = & -\,\left[\, k^{\2}+2\,m_{\1}
\left(E_{\2}+m_{\2} \right) \right]\,/\,\left\{\,4\,
\Delta\,E_{\2} \left(E_{\2}-m_{\2}\right)
 \,\left[\,\mu^{\2}+
2\,m_{\2}\left(E_{\2}-m_{\2}\right)\,
\right]^{\2}\right\}\,\,,\nonumber \\
R_{\0}^{\times} & = & 2\,(E_{\0}+2\, m_{\1})(E_{\0}+2\, m_{\2})
\left[\left(E_{\0}+m_{\1}\right)\,B\,C +
\left(E_{\0}+m_{\2}\right)\,A_{\times}\,C -
A_{\times}\,B\,\right]\,/\,\left[\,A_{\times}^{\2}B^{\2}C^{\3}\,\right]
\nonumber \\
 &  & +\,2\,\left(E_{\0}+W\right)\,/\,\left[\,A_{\times}\,B\,C^{\2}\,
 \right]\,\,,
\end{eqnarray}
with $A_{\times}=-\,\left(2\,E_{\0}\,m_{\1} + \mu^{\2}\right)$. It
is to be noted, and can be used in calculation, that the residues
of the box and crossed  residues are related by
\begin{equation}
R_{\1,\2,\0}^{\times} = - \, R_{\1,\2,\0}^{\square}[m_{1}\to
-\,m_{\1}]\,\,.
\end{equation}
However, care must be used when applying this symmetry because for
example $\sqrt{m^{\2}_{\1}}+m_{\1}=2\,m_{\1}$, while, under
$m_{\1} \to -\,m_{\1}$, $\sqrt{(-m_{\1})^{\2}}-m_{\1}=0\neq -
\,2\,m_{\1}$. The rule of thumb is that square root factors should
be left as such before applying such symmetries.

Before passing to the actual calculation of the small and large
$\mu/ m$ results, we must discuss two important technical
questions. The first is the question of the convergence of the $k$
integrals. The second is the feature of real pole contributions in
some of these residue integrals.

\subsection*{$\bullet$ Convergence.}

Individually, the leading residues terms yield divergent
integrals, both linear and logarithmic. This was not the case for
the scalar model\cite{GROSS}. However, when summed, the
divergences cancel, specifically in the limit $k \to \infty$,
\[
\begin{array}{lclrcl}
& & &16\,m_{\1}^{\2}\,m_{\2}^{\2}\,k^{\2}\, R_{\1}^{\square} & = &
-\,\displaystyle{\frac{m_{\2}^{\2}}{W}}
-\,\displaystyle{\frac{(\mu^{\2}+2\,m_{\1}m_{\2}-3\,m_{\1}^{\2})m_{\2}^{\2}}{W\,m_{\1}\,k}}
+ \, \mbox{O}\left(\frac{1}{k^{\2}}\right)\,\,,
\\
 & & &16\,m_{\1}^{\2}\,m_{\2}^{\2}\,k^{\2}\, R_{\2}^{\square} & = &
+\,\displaystyle{\frac{m_{\1}^{\2}}{W}} -
\,\displaystyle{\frac{(\mu^{\2}+2\,
m_{\1}m_{\2}-3\,m_{\2}^{\2})m_{\1}^{\2}}{W\,m_{\2}\,k}} + \,
\mbox{O}\left(\frac{1}{k^{\2}}\right)\,\,,\\
& & & 16\,m_{\1}^{\2}\,m_{\2}^{\2}\,k^{\2}\,R_{\0}^{\square} & = &
+\,\Delta  + \displaystyle{\frac{(\mu^{\2}+2\,
m_{\1}m_{\2})(m_{\1}^{\3}+m_{\2}^{\3})/W-3\,m_{\1}^{\2}m_{\2}^{\2}}{m_{\1}\,m_{\2}\,k}}
+ \, \mbox{O}\left(\frac{1}{k^{\2}}\right)\,\,.
\end{array}
\]
Consequently,
\[
16\,m_{\1}^{\2}\,m_{\2}^{\2}\,k^{\2}\, R^{\square}
=-\,\displaystyle{\frac{m_{\1}^{\2}m_{\2}^{\2}}{2\,k^{\3}}} +
\displaystyle{\mbox{O}\left(\frac{1}{k^{\5}}\right)}\,\,\,\,\,\mbox{and}\,\,\,\,\,
16\,m_{\1}^{\2}\,m_{\2}^{\2}\,k^{\2}\, R^{\times}=
+\,\displaystyle{\frac{m_{\1}^{\2}m_{\2}^{\2}}{2\,k^{\3}}} +
\displaystyle{\mbox{O}\left(\frac{1}{k^{\5}}\right)}\,\,.
\]
Both these results lead to convergent integrals, however, when
summed, the leading terms again cancel and finally
\begin{equation}
16\,m_{\1}^{\2}\,m_{\2}^{\2}\,k^{\2}\, (R^{\square}+R^{\times}) =
-\,\displaystyle{\frac{63\,m_{\1}^{\3}m_{\2}^{\3}}{4\,k^{\5}}} +
\displaystyle{\mbox{O}\left(\frac{1}{k^{\7}}\right)}\,\,,
\end{equation}
which is a highly convergent integrand. Notice that this leading
order result is symmetric under $m_{\1} \leftrightarrow m_{\2}$.
We have not specified which mass, $m_{\1}$ or $m_{\2}$, is the
lower mass and the Feynman diagrams are clearly symmetric under
the interchange $m_{\1} \leftrightarrow m_{\2}$. {\em Any final
results must therefore be symmetric under this symmetry}. This
feature may be used as a test of all of the following results.

\subsection*{$\bullet$ Poles.}

By explicit observation the quadratic term in the denominator of
$R_{\1}^{\square}$ vanishes at $\mu^{\2}=2\,
m_{\1}\,(E_{\1}+m_{\1})$. Poles also occur in the expression for
$R_{\0}^{\square}$ when $A_{\square}=0$, i.e. at $\mu^{\2}=2
m_{\1}\,E_{\0}$. Both of these conditions correspond to the {\em
same} value of $k$, which we indicate by $k_p$,
\begin{equation}
k_p^{\2} =\left(\frac{\mu^{\2}}{2\,m_{\1}}\right)^{\2} - \mu^{\2}
\end{equation}
No other residues have poles. Thus, $R_{\1}^{\square}$ and
$R_{\0}^{\square}$ exhibit double and single poles on the real
axis at $k_p$. However, when summed {\em all pole contributions
cancel}. This is demonstrated in some detail in the Appendix. The
cancellation of the double pole is simple to show. That of the
single pole which receives a contribution from $R_{\1}^{\square}$
and four contributions  from $R_{\0}^{\square}$, one from each
term in the last line of Eq.(\ref{residues}), is more cumbersome
to see. However, it must be proved since it would otherwise
dominate the large $\mu/m$ calculation, and radically change our
conclusions.

\section*{III. THE EXCHANGE OF SMALL MASS SCALARS}

For incoming fermions with mass $m_{\1}$ and $m_{\2}$ interacting
by the exchange of a  scalar with mass $\mu \ll m_{\1,\2}$,
$R^{\square}$ and  $R^{\times}$ contribute to the invariant
scattering amplitude only for value of $k\ll m_{\1,\2}$ (indeed of
the order of $\mu$). In this small $\mu$ limit, we may use the
approximation
\[E_{\1,\2}=\sqrt{k^{^{2}}+m_{\1,\2}^{^{2}}}\approx
\, m_{\1,\2}+\frac{k^{\2}}{2\,m_{\1,\2}}\,\,.
\]
We note, as an aside that for small $\mu$ ($\ll m_{\1,\2}$) there
are no poles on the real axis. Now it is easy to show that
\[ R_{\1}^{\square}/R_{\2}^{\square} = \mbox{O}[(\mu/m)^{\8}]\ll 1\,\,. \]
Whence in the rest of this Section $R_{\1}^{\square}$ will be
neglected. The other residue contributions yield
\begin{eqnarray*}
k^{\2}\,R_{\2}^{\square} & \approx &
-\,\frac{2\,m_{\1}m_{\2}}{W}\,\frac{1}{E_{\0}^{^4}}\,+\,
\frac{1}{2\,W}\,\frac{k^{\2}}{E_{\0}^{^4}}\,-\,
\frac{m_{\1}}{m_{\2}\,W}\,\frac{k^{\4}}{E_{\0}^{^6}}
 \,\,,\\
k^{\2}\,R_{\2}^{\times} & \approx &
-\,\frac{2\,m_{\1}m_{\2}}{\Delta}\,\frac{1}{E_{\0}^{^4}}\,-\,
\frac{1}{2\,\Delta}\,\frac{k^{\2}}{E_{\0}^{^4}}\,-\,
\frac{m_{\1}}{m_{\2}\,\Delta}\,\frac{k^{\4}}{E_{\0}^{^6}}
 \,\,, \\
k^{\2}\,R_{\1}^{\times} & \approx &
+\,\frac{2\,m_{\1}m_{\2}}{\Delta}\,\frac{1}{E_{\0}^{^4}}\,+\,
\frac{1}{2\,\Delta}\,\frac{k^{\2}}{E_{\0}^{^4}}\,+\,
\frac{m_{\2}}{m_{\1}\,\Delta}\,\frac{k^{\4}}{E_{\0}^{^6}}
 \,\,,  \\
k^{\2}\,R_{\0}^{\square} & \approx &
+\,\frac{3}{4}\,\frac{k^{\2}}{E_{\0}^{^5}}\,-\,
\frac{\Delta}{2\,m_{\1}m_{\2}}\,\frac{k^{\2}}{E_{\0}^{^4}}\,+\,
\frac{\mu^{\2}\Delta}{2\,m_{\1}\,m_{\2}}\,\frac{k^{\2}}{E_{\0}^{^6}}
 \,\,, \\
k^{\2}\,R_{\0}^{\times} & \approx &
-\,\frac{3}{4}\,\frac{k^{\2}}{E_{\0}^{^5}}\,-\,
\frac{W}{2\,m_{\1}m_{\2}}\,\frac{k^{\2}}{E_{\0}^{^4}}\,+\,
\frac{\mu^{\2}W}{2\,m_{\1}\,m_{\2}}\,\frac{k^{\2}}{E_{\0}^{^6}}
 \,\,.
\end{eqnarray*}
Thus,
\begin{equation}
k^{\2}\, \left[ R^{\square}(k) + R^{\times}(k)\,\right] \approx
-\,2\,m\,\frac{1}{E_{\0}^{^4}}\,+\,
\left(\,\frac{1}{2\,W}\,-\,\frac{1}{m_{\1}}\,\right)\,\frac{k^{\2}}{E_{\0}^{^4}}\,+\,
\left(\,\frac{1}{W}\,+\,\frac{1}{m_{\1}}\,\right)\,\frac{k^{\4}}{E_{\0}^{^6}}\,+\,
\frac{\mu^{\2}}{m_{\1}}\,\frac{k^{\2}}{E_{\0}^{^6}}\,\,,
\end{equation}
and, by making use of the elementary integrals
\[ \frac{4\,\mu^{\3}}{\pi}\,\int_{\0}^{\infty} \frac{\mbox{d}k}{E_{\0}^{^{4}}}
=\,\frac{4\,\mu}{\pi}\,\int_{\0}^{\infty} \frac{k^{\2}
\mbox{d}k}{E_{\0}^{^{4}}} =
\,\frac{16\,\mu}{3\,\pi}\,\int_{\0}^{\infty} \frac{k^{\4}
\mbox{d}k}{E_{\0}^{^{6}}} =
\,\frac{16\,\mu^{\3}}{\pi}\,\int_{\0}^{\infty} \frac{k^{\2}
\mbox{d}k}{E_{\0}^{^{4}}} = 1\,\,,
\]
we find that
\begin{equation}
\mathcal{M}^{\square} +  \mathcal{M}^{\times}  \approx 2\,
m_{\1}m_{\2}\,\left( \frac{g_{\1}g_{\2}}{\pi} \right)^{^{2}}\,
\left(\, -\,\frac{\pi}{2}\,\frac{m}{\mu^{\3}} \,
+\,\frac{5\,\pi}{16}\,\frac{1}{W\,\mu}\,\right) \,\,.
\end{equation}
Comparing now this fourth-order total scattering amplitude,
 \begin{equation}
\mathcal{M}^{\square}+\mathcal{M}^{\times} \approx - \,
\frac{g_{\1}^{^{\2}}g_{\2}^{^{\2}}}{\pi}\,
\frac{m_{\1}^{\2}m_{\2}^{\2}}{W\,\mu^{\3}}\,\left(\,1
\,-\,\frac{5}{8}\,\frac{\mu^{\2}}{m_{\1}m_{\2}}\,\right)\,\,,
\end{equation}
with the one boson exchange amplitude (tree diagram)
\begin{equation}
-\,4\,m_{\1}m_{\2}\,\frac{g_{\1}g_{\2}}{\mu^{\2}}\,\, ,
\end{equation}
 we find that the fourth-order amplitude is greater or
comparable to the second-order amplitude when
\begin{equation}
\label{condsmall} \frac{g_{\1}g_{\2}}{4\,\pi} \geq
\frac{\mu}{m}\,\left(\,1
\,+\,\frac{5}{8}\,\frac{\mu^{\2}}{m_{\1}m_{\2}}\,\right)\,\,,
\end{equation}
which, to the leading order, reproduces exactly the Hulthen
inequality, where $g^{\2}_{\eff}=g_{\1}g_{\2}$. We have explicitly
calculated and exhibited the correction term in the above
inequality, and we will refer to this factor in our conclusions.

\section*{IV. THE EXCHANGE OF HIGH MASS SCALARS}

The high $\mu/m$ limit is more difficult to treat and we rely upon
numerical tests of the following expressions. We have three masses
in our calculation of $\mathcal{M}^{\square,\times}$ so if we
consider an adimensional expression, it can only be a function of
$\mu/m_{\1}$ and $\mu/m_{\2}$ or alternatively of
\[
\omega=\frac{\mu^{\2}}{m_{\1}m_{\2}}\,\,\,\,\,\,\,\mbox{and}\,\,\,\,\,\,\,
\rho=\frac{m_{\1}}{m_{\2}}\,\,.
\]
Indeed,
\begin{equation}
-\,\mu^{\2}\,\int_{\0}^{\infty}\mbox{d}k \,\,k^{\2}\,\left[
R^{\square}(k)+R^{\times}(k)\right] =
F\left(\omega\,,\,\rho\right)\,\,,
\end{equation}
and this can be tested numerically. Now we try to parameterize
$\mathcal{M}^{\square,\times}$ by a form derived in the scalar
model case. We write,
\begin{equation}
-\,\mu^{\2}\,\int_{\0}^{\infty}\mbox{d}k \,\,k^{\2}\,\left[
R^{\square}(k)+R^{\times}(k)\right] = \frac{\alpha}{\omega}\,
\left( \,\ln\omega + \frac{1+\rho^{\2}}{1-\rho^{\2}} \ln
\rho\,\right)\,\,.
\end{equation}
The value $\alpha=1$ reproduces the scalar model result. This
phenomenological form has been tested for a wide but limited range
of $\omega$ and $\rho$ values, specifically for
\[
\omega=10^{\6}\,,\,10^{\7}\,,\,10^{\8}\,\,\,\,\,\,\,
\mbox{and}\,\,\,\,\,\,\,\rho=2\,,\,10\,,\,50\,\,.
\]
In the following Table
\[
\begin{array}{|c|c|c|}
\hline \,\,\,\omega\,\,\, & \,\,\,\rho\,\,\, &
\mbox{\,\,\,Phen/Num\,\,\,}
\\ \hline \hline
10^{\6} & \,\,\,2 & .995 \\
\hline
10^{\6} & 10 & .989 \\
\hline
10^{\6} & 50 & .978 \\
\hline \hline
10^{\7} & \,\,\,2 & 1.005\,\, \\
\hline
10^{\7} & 10 & 1.000\,\, \\
\hline
10^{\7} & 50 & .993 \\
\hline \hline
10^{\8} & \,\,\,2 & 1.012\,\, \\
\hline
10^{\8} & 10 & 1.008\,\, \\
\hline
10^{\8} & 50 & 1.003\,\, \\
\hline
\end{array}
\]
we give the comparison of phenomenological/numerical (Phen/Num)
results for a best fit value of $\alpha$,
\begin{equation}
\alpha=0.663\,\,.
\end{equation}
We see that to within a few per cent the agreement is good. We
could of course improve the comparison if we included a constant
term $\ln \beta$ in the brackets which could correspond to e
renormalization of the logarithmic terms. However, we consider
this an excessive finess. The important point is that the large
$\mu/m$ behavior is similar to the scalar model result. The high
$\mu$ resonance inequality thus reads
\begin{equation}
\label{condition} \frac{g^{\2}_{\eff}}{2\,\pi^{^{\2}}}  \geq
\frac{\mu^{\2}}{\alpha\, m_{\1}m_{\2}} \,\mbox{\large /}\,\left(
\,\ln\frac{\mu^{\2}}{m_{\1}m_{\2}} +
\frac{1+\rho^{\2}}{1-\rho^{\2}} \ln \rho\,\right)\,\,.
\end{equation}

\section*{V. CONCLUSIONS}

We have applied in this paper a field theoretic approach to the
determination of the coupling strengths needed for the existence
of a fermion-fermion bound state via scalar boson exchanges. For
low $\mu/m$, we again find the Hulthen inequality\cite{FLU,GROSS}
as seen in the scalar model. For high $\mu / m$, we obtain an even
more restrictive condition (\ref{condition}), a result again
similar to the scalar field model\cite{EPJC}. The similarity
between the scalar field model and this calculation suggests that
the bound state inequality condition depends essentially upon the
exchanged particles rather than the incoming ones. This was by no
means obvious since the numerators of the residues are different
in the two cases. Indeed at first sight the fermion-fermion model
seemed to yield divergent results as a simple power count of the
$\boldsymbol{k}$-integral suggests. We have shown in this paper
that the individual divergence contributions cancel. We have also
shown that the real pole contributions to $\mathcal{M}^{\square}$
also cancel both for the double and single poles. Again it is not
clear if this would happen with say vector particle exchanges and
it must be said that a contribution from a simple pole would
completely alter our high $\mu / m$ results. For the existence of
a relativistic bound state such a contribution could even be
desirable.

There is however a problem with our results for small $\mu /m$ and
the arguments based upon the relativistic corrections to the
Schr\"odinger equation mentioned in the introduction. The Dirac
equation with a scalar potential contains a Darwin term as does
the better known electrostatic case\cite{SAK}. This lead us to
expect, at least for small $\mu / m$ (nonrelativistic) a coupling
amplification. We have purposefully kept the
$\mbox{O}(\mu^{\2}/m^{\2})$ corrections in the small $\mu / m$
case and as can be seen in the result (\ref{condsmall}) the
corrections terms correspond to a coupling deamplification. The
coupling constants must be somewhat {\em increased} to compensate
the correction terms. This result is  consistent with the tougher
large $\mu / m$ inequality. We predict that the Hulthen inequality
is a lower limit inequality for any $\mu /m$. Is this disagreement
between our field theory calculation and the nonrelativistic
reduced mass equation serious? This may well be a matter of
opinion but some
observations are in order:\\

- The Hulthen inequality is {\em not} exactly in agreement with
the Yukawa numeric inequality. So, we have a formal discrepancy
even
neglecting the relativistic correction terms;\\

- The higher order Feynman diagrams cannot be parameterized by a
simple Yukawa potential. However, the Coulomb potential works
admirably well for Hydrogen like atoms except for one of the
supreme successes of field theory, the Lamb shift. Unfortunately,
we known of no direct way to derive the potential bound state
spectrum from field theory;\\

- It must also be remembered that not all the fourth order Feynman
diagrams have been calculated.\\

\noindent Nevertheless, we remain troubled by this result. At the
very least, we must moderate any expectations for a weak
interaction calculation in which intermediate vector particles are
exchanged. We expect the same low $\mu /m$ inequality (except
perhaps for the correction term) but hope for a very different
high $\mu / m$ result.

Our results have one physical consequence, we predict that weak
interacting fermion-fermion (or scalar-scalar) particles cannot
produce a bound state simply by Higgs boson exchanges\cite{PDG}.
It is our intention to tackle the full weak interaction case in
the near future.

\newpage
\section*{APPENDIX: THE POLE CONTRIBUTIONS}

In this Appendix, we calculate the pole contributions of
$R_{\1}^{\square}$ and $R_{\0}^{\square}$ at
\[
k_p =\sqrt{\left(\frac{\mu^{\2}}{2\,m_{\1}}\right)^{\2} -
\mu^{\2}}\,\,.
\]
For convenience, we define the functions $F(k)$, $G(k)$ and
$\boldsymbol{H}(k)$ by
\begin{eqnarray}
k^{\2}R_{\1}^{\square}(k) & = & F(k)\,\,,\\
k^{\2}R_{\0}^{\square}(k) & = &
G(k)\,+\,\sum_{n=\1}^{\3}H_n(k)\,\,,
\end{eqnarray}
where
\begin{eqnarray*}
G(k) & = & 2\,k^{\2}\,(E_{\0}+2\, m_{\2})(2\,m_{\1}-E_{\0})
\left(E_{\0}-m_{\1}\right)\,/\,
\left(\,A_{\square}^{\2}\,B\,\,C^{\,\2}\,\right)\,\,,\\
H_{\1}(k) & = &2\,k^{\2}\,(E_{\0}+2\, m_{\2})(2\,m_{\1}-E_{\0})
\left(E_{\0}+m_{\2}\right)\,/\,
\left(\,A_{\square}\,B^{\2}C^{\,\2}\,\right)\,\,, \\
H_{\2}(k) & = &-\, 2\,k^{\2}\, (E_{\0}+2\,
m_{\2})(2\,m_{\1}-E_{\0})\,/\,
\left(\,A_{\square}\,B\,\,C^{\,\3}\,\right)\,\,, \\
H_{\3}(k) & = &2\,k^{\2}\,\left(m_{\1}- m_{\2} -
E_{\0}\right)\,/\,\left(\,A_{\square}\,B\,\,C^{\,\2}\,\right)\,\,.
\end{eqnarray*}
The first pole terms  in the MacLaurin series of these functions
are
\[
\left\{\, F(k)\,,\,G(k)\,,\,\boldsymbol{H}(k) \,\right\} =
\left\{\, \frac{F^{^{\m2}}(k_{p})}{(k-k_{p})^{^{2}}}\,+\,
\frac{F^{^{\min1}}(k_{p})}{k-k_{p}}\,,\,
\frac{G^{^{\m2}}(k_{p})}{(k-k_{p})^{^{2}}}\,+\,
\frac{G^{^{\min1}}(k_{p})}{k-k_{p}}\,,\,
\frac{\boldsymbol{H}^{^{\min1}}(k_{p})}{k-k_{p}}\,\right\}\, +\,
\mbox{O}\left(1\right)\,\,,
\]
where $F^{^{\m2}}(k_{p})$ is the coefficient of $(k-k_{p})^{^{\mi
\2}}$ in $F(k)$ and so forth.

Now for the double pole, we find that the only two contributions
are
\begin{equation}
F^{^{\m2}}(k_{p})=-\,G^{^{\m2}}(k_{p})=
\frac{(2\,m_{\1}^{\2}-\mu^{\2})(4\,m_{\1}m_{\2}+\mu^{\2})}{16\,m_{\1}^{\2}W\,\mu^{\4}}
\,\,k_{p}^{\2}\,\,,
\end{equation}
whence their sum cancels.

The single pole contributions can be written as
\begin{equation}
\label{polec} \left\{\, F^{^{\min1}}(k_{p})\,,\,
G^{^{\min1}}(k_{p})\,,\,
\boldsymbol{H}^{^{\min1}}(k_{p})\,\right\} =
\frac{k_{p}}{32\,m_{\1}^{\2}W^{\2}\,\mu^{\6}}\,\left\{\,
f^{^{\min1}}_{p}\,,\, g^{^{\min1}}_{p}\,,\, \boldsymbol{h}^{^{
\min1}}_{p}\,\right\}\,\,,
\end{equation}
and in the Table we list the factors in graph brackets above as a
series in even powers of $\mu$. For example,
\[
f^{^{\min1}}_{p} = -\,2\,W\,\mu^{\6} -(m_{\2}+\Delta)\,W\,\frac{\mu^{\4}}{2\,m_{\1}}
-\, m_{\2}W\,\frac{\mu^{\2}}{8\,m_{\1}^{^{3}}}\,\,.
\]
The important point is contained in the last line of the Table.
All single pole contributions also cancel. Thus, in conclusion,
there are no real axis poles in
$k^{\2}\left(R^{\square}+R^{\times}\right)$.

\newpage
\[
\begin{array}{l|c|c|c|c}
 & \mu^{\6} & \mu^{\4}\,/\,2\,m_{\1} & \mu^{\2}\,/ \, 8\,m_{\1}^{\3} &
 \mu^{\0}\,/\,3\,m_{\1}^{\5}m_{\2} \\
\hspace*{1cm} & \hspace*{2.6cm} & \hspace*{3.8cm} &
\hspace*{2.6cm} & \hspace*{2.6cm}
\\ \hline
 & & & & \\ f_{p}^{^{\min1}} & -2\,W&-(m_{\2}+\Delta)W &-m_{\2}W & 0 \\
& & & &  \\ \hline & & & & \\ g_{p}^{^{\min1}} & m_{\1}+2\,W
&2\,\Delta W - 2\,m_{\1}^{\2}+3\,m_{\1}m_{\2} &
2\,m_{\2}\Delta  &  - (m_{\2}+W)\\
& & & &  \\ \hline & & & & \\ h_{p,\1}^{^{\min1}} & - m_{\1} &
2\,m_{\1}^{\2} - 3\,m_{\1}m_{\2}  &
m_{\2} (2\,m_{\1} - \Delta) & m_{\2}\\
& & & &  \\ \hline
& & & & \\ h_{p,\2}^{^{\min1}} & 0 & - m_{\1} W & - \Delta W & W  \\
& & & &  \\ \hline
& & & & \\ h_{p,\3}^{^{\min1}} & 0 & 2\,m_{\1} W & \Delta W & 0 \\
& & & &  \\ \hline \hline
& & & & \\ Sum & 0 & 0 & 0 & 0 \\
& & & &  \\ \hline
\end{array}
\]
{\bf Table.} The coefficients of powers of $\mu^{\2}$ in the
factors $f^{^{\min1}}_{p}$, $g^{^{\min1}}_{p}$ and
$\boldsymbol{h}^{^{ \min1}}_{p}$ for the single pole
contributions. These factors are defined in Eq.(\ref{polec}).

\newpage

\begin{figure}[hbp]
\hspace*{-2.5cm}
\includegraphics[width=19cm, height=24cm, angle=0]{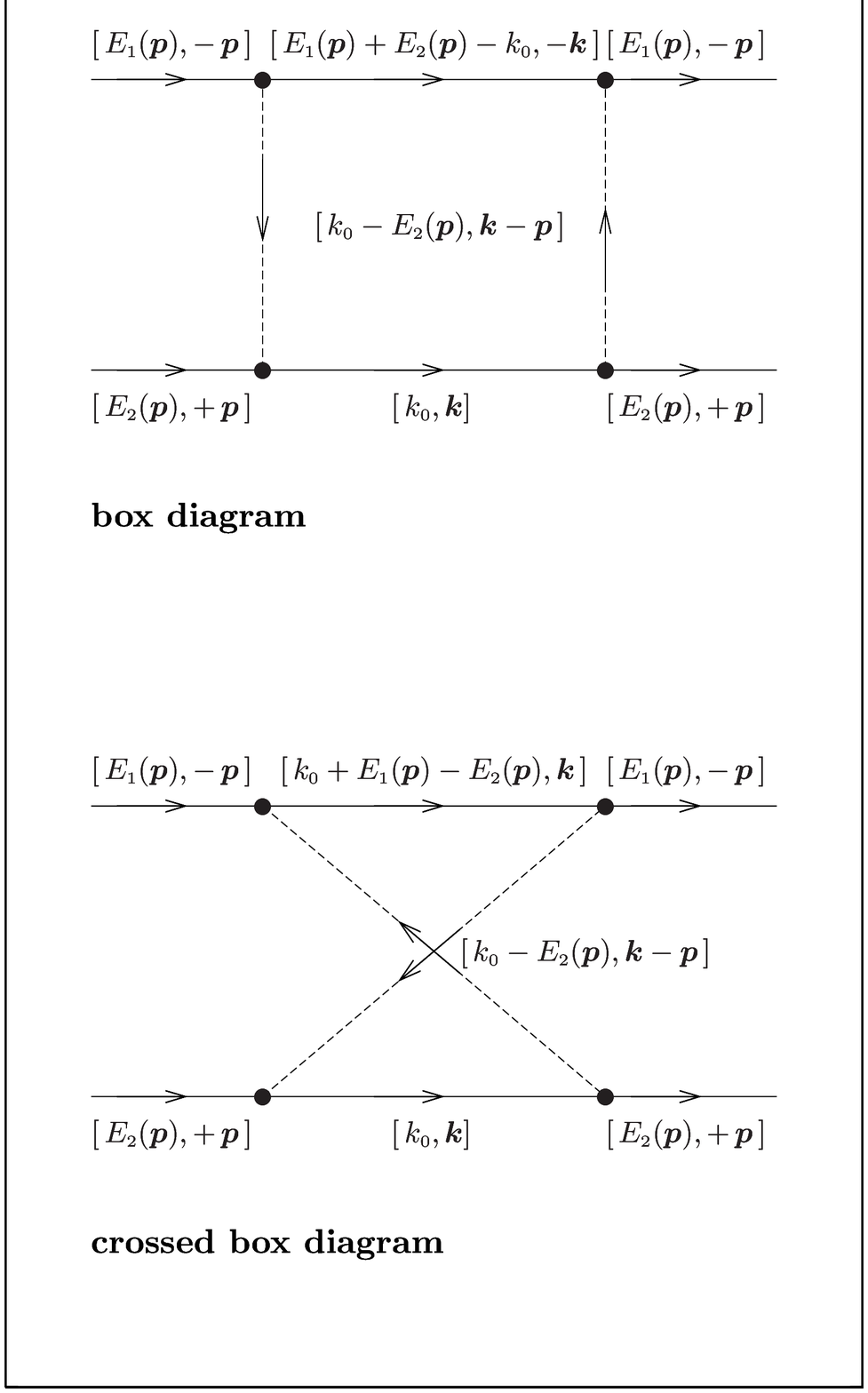}
\vspace*{-2cm}
 \caption{The fourth order box and crossed box diagrams in a ferminonic field
 model evaluated in the center of mass frame for scattering in the forward
direction.}
\end{figure}


\begin{thebibliography}{99}

\bibitem{PDG} W.M. Yao et al., "Review of particle physics", J.
Phys. G {\bf 33}, 156-164 (2006).

\bibitem{COHEN}
C. Cohen-Tannoudji, B. Diu and F. Lalo\"e, {\em Quantum
mechanics}, John Wiley \& Sons, Paris (1977).


\bibitem{FLU}
S. Fl\"ugge, {\em Practical quantum mechanics}, Springer-Verlag,
Berlin (1999).


\bibitem{ZUB}
C. Itzykson and J.B. Zuber, {\em Quantum field theory}, McGraw-Hill,
Singapore (1985).


\bibitem{SAK}
J.\,J. Sakurai, {\em Advanced quantum mechanics}, Addison-Wesley, New York
(1987).


\bibitem{AYC}
S. De Leo and P. Rotelli, "Amplification of coupling for Yukawa
potentials", Phys. Rev. D {\bf 69}, 034006-5 (2004).

\bibitem{G69}
F. Gross, "Three-dimensional covariant integral equations for
low-energy systems", Phys. Rev. {\bf 186}, 1448-1462 (1969)


\bibitem{GROSS}
F. Gross, {\em Relativistic quantum mechanics and field theory}, John \&
Wiley Sons, New York (1993).

\bibitem{EPJC}
S. De Leo and P. Rotelli, "Bound state inequality for high mass
exchanges in a scalar field model", to appear in Eur. Phys. J. C
(2008).


\end{thebibliography}
\end{document}